\documentclass[conference]{IEEEtran}
\IEEEoverridecommandlockouts
\usepackage{amsmath,amssymb,amsfonts}
\usepackage{graphicx}
\usepackage{textcomp}
\usepackage{xcolor}
\def\BibTeX{{\rm B\kern-.05em{\sc i\kern-.025em b}\kern-.08em
    T\kern-.1667em\lower.7ex\hbox{E}\kern-.125emX}}

\usepackage[english]{babel}
\usepackage[utf8]{inputenc}

\usepackage{algorithm}   
\usepackage{algpseudocode}
\usepackage{float}
\usepackage{lipsum}
\usepackage{hyperref}
\usepackage{multirow}
\usepackage{amsthm}
\usepackage{array}
\usepackage{scalerel}
\usepackage{bm}
\usepackage{tikz}
\usepackage{csquotes}

\usetikzlibrary{svg.path}
\usetikzlibrary{calc, fit}

\usepackage[backend=biber, bibstyle=ieee,citestyle=numeric-comp, sorting=nyt, dashed=false]{biblatex}
\theoremstyle{remark}

\theoremstyle{definition}

\newcolumntype{M}[1]{>{\centering\arraybackslash}m{#1}}

\newcolumntype{L}[1]{>{\raggedright\arraybackslash}m{#1}}

\newcommand{\algstep}[1]{\item[]\medskip\hrule\kern 2pt\hbox to \linewidth{\hspace{\labelsep}\textbf{#1}\hfill}\hrule}

\newlength\myindent
\makeatletter
\newlength{\trianglerightwidth}
\algnewcommand{\LineCommentContAfter}[1]{\Statex \hskip\ALG@tlm%
    \parbox[t]{\dimexpr\linewidth-\ALG@tlm}{\hangindent=\trianglerightwidth \hangafter=1 \strut$\triangleright$ #1\strut}}
\algnewcommand{\LineCommentContBefore}[1]{\Statex \hskip\ALG@thistlm%
    \parbox[t]{\dimexpr\linewidth-\ALG@thistlm}{\hangindent=\trianglerightwidth \hangafter=1 \strut$\triangleright$ #1\strut}}
\makeatother



\makeatletter
\algnewcommand{\NewLineCode}[1]{\Statex \hskip\ALG@thistlm#1}
\algnewcommand{\LineCodeCont}[1]{\State 
    \parbox[t]{\dimexpr\linewidth-\ALG@thistlm}{\hangindent=\trianglerightwidth \hangafter=1 \strut #1\strut}}
\makeatother

\definecolor{orcidlogocol}{HTML}{A6CE39}
\tikzset{
  orcidlogo/.pic={
    \fill[orcidlogocol] svg{M256,128c0,70.7-57.3,128-128,128C57.3,256,0,198.7,0,128C0,57.3,57.3,0,128,0C198.7,0,256,57.3,256,128z};
    \fill[white] svg{M86.3,186.2H70.9V79.1h15.4v48.4V186.2z}
                 svg{M108.9,79.1h41.6c39.6,0,57,28.3,57,53.6c0,27.5-21.5,53.6-56.8,53.6h-41.8V79.1z M124.3,172.4h24.5c34.9,0,42.9-26.5,42.9-39.7c0-21.5-13.7-39.7-43.7-39.7h-23.7V172.4z}
                 svg{M88.7,56.8c0,5.5-4.5,10.1-10.1,10.1c-5.6,0-10.1-4.6-10.1-10.1c0-5.6,4.5-10.1,10.1-10.1C84.2,46.7,88.7,51.3,88.7,56.8z};
  }
}

\newcommand\orcidicon[1]{\href{https://orcid.org/#1}{\mbox{\scalerel*{
\begin{tikzpicture}[yscale=-1,transform shape]
\pic{orcidlogo};
\end{tikzpicture}
}{|}}}}

\begin{document}

\title{Data Privacy in Multi-Cloud: An Enhanced Data Fragmentation Framework\\
\thanks{}
}

\author{\IEEEauthorblockN{Randolph Loh}
\IEEEauthorblockA{\textit{Cyber Security Strategic Technology Centre} \\
\textit{Singapore Technologies Engineering Ltd}\\
Singapore \\
\orcidicon{0000-0001-8132-4266} \href{https://orcid.org/0000-0001-8132-4266}{0000-0001-8132-4266}}
\and
\IEEEauthorblockN{Vrizlynn L. L. Thing}
\IEEEauthorblockA{\textit{Cyber Security Strategic Technology Centre} \\
\textit{Singapore Technologies Engineering Ltd}\\
Singapore \\
\orcidicon{0000-0003-4424-8596} \href{https://orcid.org/0000-0003-4424-8596}{0000-0003-4424-8596}}
}

\maketitle


\begin{abstract}
Data splitting preserves privacy by partitioning data into various fragments to be stored remotely and shared. It supports most data operations because data can be stored in clear as opposed to methods that rely on cryptography. However, majority of existing data splitting techniques do not consider data already in the multi-cloud. This leads to unnecessary use of resources to re-split data into fragments.
This work proposes a data splitting framework that leverages on existing data in the multi-cloud. It improves data splitting mechanisms by reducing the number of splitting operations and resulting fragments. Therefore, decreasing the number of storage locations a data owner manages. Broadcasts queries locate third-party data fragments to avoid costly operations when splitting data. This work examines considerations for the use of third-party fragments and application to existing data splitting techniques. The proposed framework was also applied to an existing data splitting mechanism to complement its capabilities.

\end{abstract}


\section{Introduction}\label{sect:intro}
Data privacy is a concern when storing data in the Cloud as it potentially gives external entities access to the data. Extensive research on protecting the privacy of outsourced data resulted in a plethora of data protection technologies seeking to safely outsource sensitive data to the Cloud \cite{domingo2019privacy}.
The multi-cloud is associated with public cloud solutions as a heterogeneous collection of multiple CSPs with multi-tier applications that migrated from private systems attract several user archetypes, while the private cloud restricts its services to a selective class of users \cite{ferry2013towards, petcu2013multi, jamshidi2015cloud, 10.5555/2385915}. In terms of data storage, users can securely outsource data of different levels of sensitivity leveraging these architectures. For example, non-sensitive data can be stored in the public cloud and sensitive data in private clouds through the hybrid cloud infrastructure. The distributed and unconnected nature of the multi-cloud prevents collusion between CSPs in a way that undermines the privacy of the data owner \cite{domingo2019privacy}. 
However, there are still concerns on security and privacy \cite{bohli2013security}.

Data splitting segments data into fragments in a way that allows it to be stored at various locations such that less sensitive fragments may be outsourced for storage in the multi-cloud. This allows data owners to partially share their data with other entities. As opposed to cryptographic techniques where data is stored encrypted \cite{domingo2019privacy}, less sensitive fragments can be stored in clear to support various operations. However, many data splitting techniques do not consider to use of pre-existing data in the multi-cloud, and consequently, waste resources repeatedly splitting data that are readily available.

\paragraph{Contributions}
This work introduces a simple yet non-trivial query step in the data splitting process. The main contributions of this work are summarised as follows:
\begin{itemize}
\item This work proposes a data splitting framework that leverages data that exist in the Cloud to improve the data splitting mechanisms. The framework tries to reduce the number of operations required to fragment data and is thus more efficient. The resulting data fragments outsourced for storage are also reduced, thereby reducing the number of CSPs a user has to manage.
\item The proposed framework exploits existing data in the Cloud to increase reliability and availability. Checks are performed during retrieval where missing or corrupted fragments can be rebuilt referencing third-party sources.
\item An analysis on the considerations and application of the proposed framework on existing data splitting techniques. The framework was also applied to the semantic data splitting mechanism described in \cite{sanchez2017privacy}.
\end{itemize}

\paragraph{Paper Organisation}
This paper is organised as follows. Data splitting is described in Section~\ref{sect:data-split}. Section~\ref{sect:data-split-leveraging} introduces the proposed framework and considerations when applying to existing data splitting techniques. Section~\ref{sect:sem-data-splitting-leveraging} details the proposed framework complementing the work of \cite{sanchez2017privacy} before concluding in Section~\ref{sect:con}.


\section{Data splitting}\label{sect:data-split}
Data splitting protects data whilst observing given privacy requirements \cite{yang2015hybrid}. Data is split at the \emph{attribute}, \emph{byte}, or \emph{semantic}-level \cite{domingo2019privacy}. Data splitting should constitute a \emph{lossless} process such that the original data can be reconstructed from its data fragments after being distributed and stored across the multi-cloud. Less sensitive data fragments can be stored in clear thus preserving some utility and therefore suitable for sharing.
Storing data fragments at different locations makes it difficult for potential attackers for it will be necessary to target multiple, if not all, relevant CSPs to retrieve the original data and extract useful information \cite{dev2012approach}. Individual fragments neither allow re-identification nor leak sensitive information as information inference is prevented \cite{domingo2019privacy}.

\subsection{System architecture}
\begin{figure}[htbp]
\centering
\includegraphics[height=5cm]{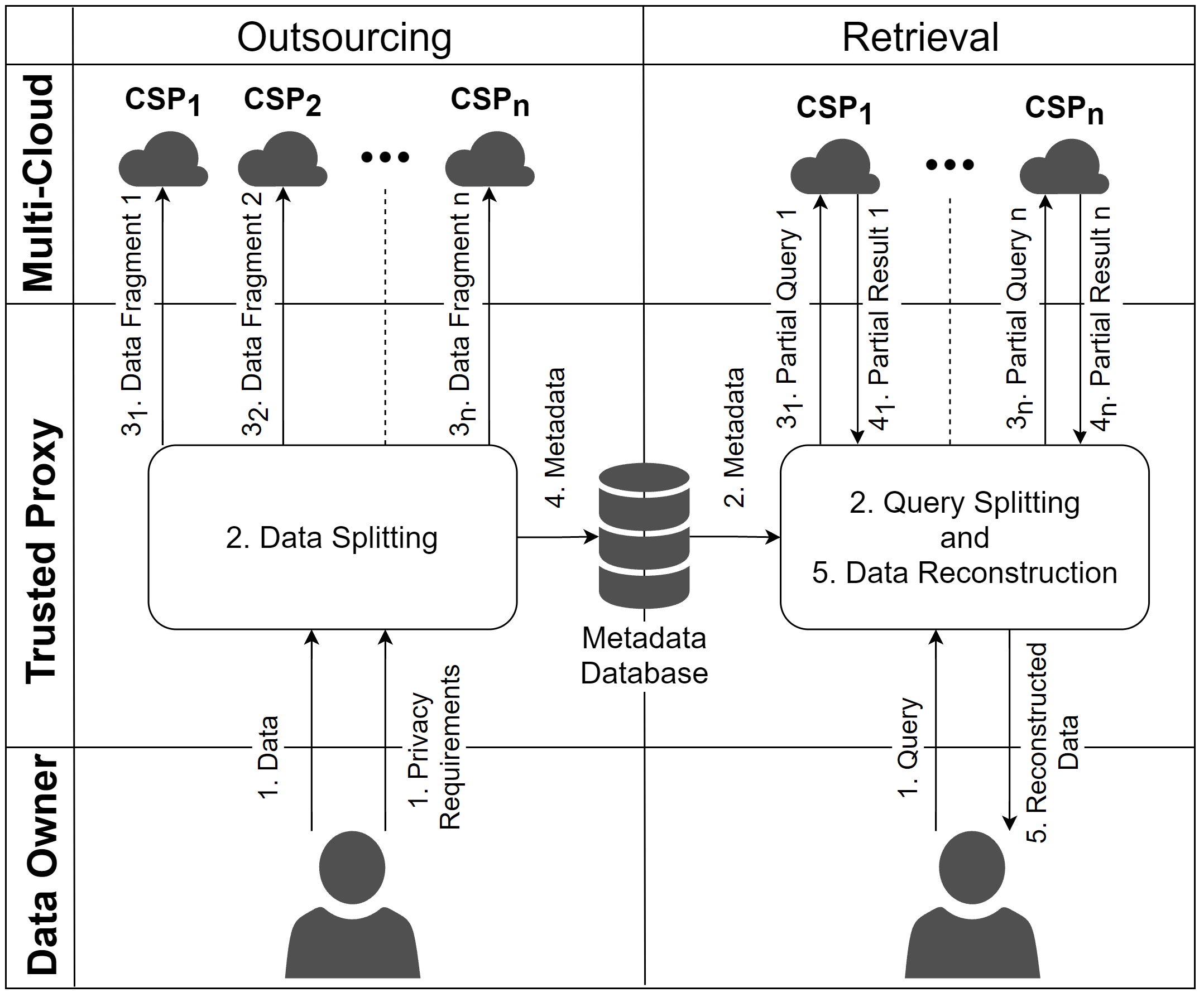}
\caption{Data outsourcing (left) and retrieval (right) through data splitting.}
\label{fig:arch1}
\end{figure}

Data splitting is namely performed between three entities: the \emph{data owner}; a \emph{trusted proxy}; and the \emph{CSP(s)} \cite{domingo2019privacy}. 
Their interactions are depicted in Figure~\ref{fig:arch1} and described as follows. \\
Data splitting and outsourcing:
\begin{enumerate}
\item The data owner sends the data and its privacy requirements to the trusted proxy.
\item The trusted proxy receives the data and privacy requirements from the data owner. It splits the data according to the privacy requirements.
\item The trusted proxy stores sensitive data fragments in a local database while outsourcing less sensitive data fragments to be stored and shared in the multi-cloud.
\item The trusted proxy records the locations for each data fragment and stores it as metadata which is used to retrieve and reconstruct the data to its original form.
\end{enumerate}
Data retrieval:
\begin{enumerate}
\item The data owner sends a query to the trusted proxy.
\item The trusted proxy retrieves the storage information from its database to craft queries accordingly. 
\item The trusted proxy sends partial queries to their CSPs.
\item CSPs responds to the trusted proxy with partial results.
\item The trusted proxy reconstructs the data from its partial results and returns it to the data owner.
\end{enumerate}

\subsection{Related work}

A distributed architecture using two untrusted servers was presented in \cite{aggarwal2005two}. Data partitioned across independent databases ensure that contents in any one database do not violate privacy. The databases are assumed to not communicate with each other. Data confidentiality is achieved by vertically partitioning data. Queries are also transformed before they are sent. However, the authors of \cite{Ciriani2007FragmentationAE,Ciriani2010CombiningFA} suggested the use of encryption to overcome the assumption where databases do not communicate to further improve privacy guarantees. Fragmentation was minimised because queries over fragmented and encrypted attributes are generally inefficient. 
The authors of \cite{gai2016security} proposed a scheme that encrypts data before distributing to different cloud storage facilities applying two algorithms that leveraged logical operations to manipulate data bits. The scheme was extended in \cite{li2017intelligent} where a parallelisable algorithm decides if a data packet requires more security from internal or external threats. 
In another work, users determine the number of data fragments before a file is split using a random pattern fragmentation algorithm and distributed to NoSQL databases \cite{santos2019enhancing}. A faster alternative to cryptographic approaches but carries some processing overheads. Also, secret sharing schemes and hashing algorithms can be used to allow multiple authorised entities to access the distributed data \cite{al2019big}.

\subsection{Motivation}
Data splitting when used with other privacy-preserving techniques provides stronger security guarantees but also degrades data utility and often requires added resources. 
This work attempts to avoid costly operations to increase the efficiency of data splitting mechanisms. Most proposed methods only considered outsourcing data to the multi-cloud for storage or sharing. To the best of the authors' knowledge, none have explored in detail the use of existing or published data in the Cloud when fragmenting data.
In the event where two data objects are fragmented in a way both result in equal data fragments, either data object can use the other's fragment to reconstruct itself. Owing to this, data owners may adopt readily available third-party data fragments instead of generating their own, reducing the need to process and store their own data fragments in the multi-cloud.


\begin{figure}[htbp]
\centering
\includegraphics[height=5cm]{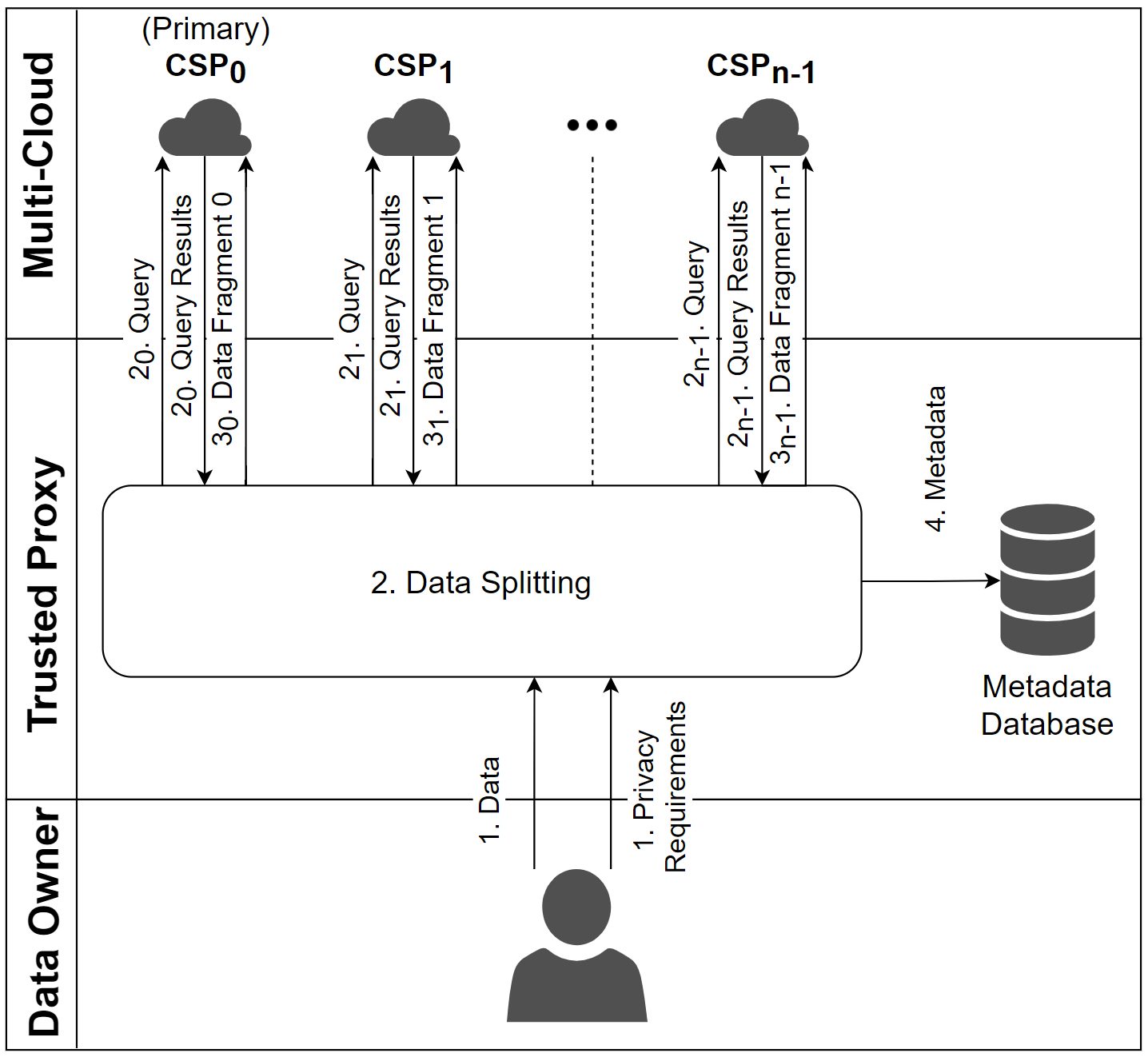}
\caption{Proposed general data splitting and storage to the multi-cloud.}
\label{fig:arch2}
\end{figure}

\section{Data splitting leveraging on existing data}\label{sect:data-split-leveraging}
Many works try to preserve privacy and utility when outsourcing data but none have been identified to take advantage of data that is readily available in the Cloud. This work argues that the data owner does not need to process and store data that already exist in the Cloud. It assumes that fragmenting various data objects will result in data fragments that are similar, if not the same, under similar data splitting policies or requirements. The proposed framework searched for these data fragments in the multi-cloud. Such data fragments hold significance because they can be used to reconstruct multiple distinct sets of data and should therefore be managed appropriately. This work considers data splitting within the context of the  multi-cloud. 

\subsection{System architecture}
Figure~\ref{fig:arch2} depicts the proposed framework involving three entities: the \emph{data owner}; a \emph{trusted proxy}; and the \emph{CSP(s)}. Here, the proxy \emph{broadcasts queries} during the splitting process. The retrieval process is unchanged.\\
Data splitting and outsourcing (proposed framework):
\begin{enumerate}
\item The data owner sends the data and its privacy requirements to the trusted proxy.
\item The trusted proxy receives the data and privacy requirements from the data owner. It \emph{broadcasts queries} to the CSPs for similar data fragments. It then splits the remaining data according to the privacy requirements. 
\item The trusted proxy stores sensitive data fragments in a local database while outsourcing less sensitive data fragments to be stored and shared in the multi-cloud.
\item The trusted proxy records the locations for each data fragment and stores it as metadata which is used to retrieve and reconstruct the data to its original form.
\end{enumerate}

\subsection{Cloud deployment models, security models, and queries}
The proposed framework is applicable to all types of Cloud deployment. Risks are subject to varying security models, level of sensitivity of data to be stored, and queries to CSPs. 

The 
\emph{Public Cloud} 
assumes CSPs are \emph{semi-honest} or \emph{honest-but-curious}. Where information may be extracted from queries and mapped to their corresponding data fragments or original data. Queries must leak as little information as possible and to use secure means communication. 
\emph{Private Clouds} 
are \emph{restricted to authorised users} and are assumed to be \emph{trusted}. Although localised queries are assumed to be secure, sending queries to remote instances retain similar concerns as queries in the public cloud.
\emph{Hybrid Clouds} 
incorporate characteristics of public and private clouds. CSPs are assumed to be \emph{semi-honest} or \emph{honest-but-curious} for higher levels of security. Data fragments may have varying sensitivity where sensitive data fragments can be stored at locations that provide more security guarantees (i.e. private cloud).
CSPs are assumed to less likely collude due to the segregation of the public and private cloud. Therefore queries can be separated accordingly, which only partially leak information. This is a stronger security guarantee compared to the public cloud.

\subsection{Primary and secondary CSPs}
Data owners can provide a list of $n$ CSPs ($list\_CSP$) during the data splitting process, the primary ($list\_CSP[0]$) and secondary CSPs ($list\_CSP[1]$ to $list\_CSP[n-1]$). 
The \emph{Primary CSP} (PCSP) is the primary location where a given data fragment will be stored. The data owner has full control over data fragments stored here. It is assumed to be the most \emph{reliable} and is the first location that is searched when queried. 
\emph{Secondary CSPs} (SCSPs) are used as alternative sources of data belonging to a mix of third-party entities. Data at these locations are assumed to be freely shared. This provides \emph{availability} through \emph{redundancy}. 
The trusted proxy broadcasts queries for each data fragment $f$ in the set $\mathbb{F}$ where each CSP will respond with the location of similar data fragments if it is found or $None$ otherwise. Essentially creating a list of $n$ locations ($sLoc$). 
The trusted proxy stores the data fragment at the PCSP if it responded $None$. 
A $|\mathbb{F}|*n$ list of storage locations ($loc\_list$) is updated accordingly. The $loc\_list$ is stored on the proxy locally and is used when retrieving data.

\begin{algorithm}
    \scriptsize
    \caption{Data outsourcing broadcast query} \label{alg:p1}
    \begin{algorithmic}[1]
        \renewcommand{\algorithmicrequire}{\textbf{Input:}}
        \renewcommand{\algorithmicensure}{\textbf{Output:}}
        \Require
        \Statex \hspace*{\algorithmicindent} $\mathbb{F}$ \Comment{\parbox[t]{.6\linewidth}{a set of data fragments}}
        \Statex \hspace*{\algorithmicindent} $list\_CSP$ \Comment{\parbox[t]{.6\linewidth}{a list of $n$ specified CSPs}}
        \Ensure
        \Statex \hspace*{\algorithmicindent} $loc\_list$ \Comment{\parbox[t]{.6\linewidth}{a $|\mathbb{F}|*n$ list of storage locations}}

        \Statex
        \For{each $f$ in $\mathbb{F}$}
            \LineCommentContAfter{broadcast to all specified CSPs querying for a similar data fragment; CSP returns the location if it exists, $None$ if it does not}
            \State $sLoc \gets$ broadcastQuery($f, list\_CSP$)
            \LineCommentContBefore{check if the primary CSP returns negative; if so, store the data fragment at the primary CSP}
            \If {not($sLoc[0]$)}
                \State  $sLoc[0] \gets$ storeFrag($f, list\_CSP[0]$)
            \EndIf
            \LineCommentContBefore{add to the storage information list}
            \State add($sLoc, loc\_list$)
        \EndFor
        \State return $loc\_list$
    \end{algorithmic}
\end{algorithm}

\subsection{Data fragment management}
Multiple data objects potentially employ the same data fragment and third-parties data fragments may be outside one's control. 
Data \emph{consistency} is an important factor in ensuring that updates and deletions are properly reflected across the multi-cloud \cite{mhaisen2020data}. Dynamic systems are necessary to upkeep countless data fragments in a distributed environment \cite{raouf2017distributed}. 
For simplicity, SCSPs are assumed to be third-parties. 

\subsubsection{Update data fragments}\label{man:up}
The proposed framework simplifies updating data fragments as only CSPs directly controlled by the data owner are affected, more specifically the data fragment that needs to be revised. There are two approaches to update data fragments. 
Approach 1 tries to replace the \emph{old} data fragment with a \emph{new} data fragment. However, old fragments may be purposed by multiple data objects. Directly replacement causes conflicts for other data objects and faults in data retrieval. 
Data fragments should only be directly replaced if no other data object is affected. 
Approach 2 straightforwardly selects a new PCSP to store the \emph{new} data fragment and records its location. 
Fragments should still adhere to their privacy requirements in both approaches. Finally, the trusted proxy should also update SCSPs storage information.



\subsubsection{Delete data fragments}\label{man:del}
Data fragments are deleted from its PCSP even though it will create conflicts for other data objects that utilise the data fragment and PCSP. 

\subsubsection{Conflicts in data fragments}\label{man:con}
Conflicts arise when a PCSP no longer contains the appropriate data fragment previously stored during the outsourcing process. This may be validated via the response of the PCSP, if the returned data fragment is consistent with SCSPs or not $None$.

\subsubsection{Resolving conflicts}\label{man:re}
The trusted proxy can query SCSPs for their respective data fragments and rebuild the appropriate data fragment when conflicts occur. Rebuilt data fragments should not be stored with the \emph{old} PCSP but a \emph{new} PCSP because the \emph{old} PCSP may have been purposed by other data objects, thus avoiding future conflicts.

\begin{algorithm}
    \scriptsize
    \caption{Data retrieval and reconstruction} \label{alg:p2}
    \begin{algorithmic}[1]
        \renewcommand{\algorithmicrequire}{\textbf{Input:}}
        \renewcommand{\algorithmicensure}{\textbf{Output:}}
        \Require
        \Statex \hspace*{\algorithmicindent} $loc\_list$ \Comment{\parbox[t]{.6\linewidth}{a $|\mathbb{F}|*n$ list of storage locations}}
        \Statex \hspace*{\algorithmicindent} $pri\_CSP_{new}$ \Comment{\parbox[t]{.6\linewidth}{a \emph{new} primary CSP (optional)}}
        \Ensure
        \Statex \hspace*{\algorithmicindent} $org\_data$ \Comment{\parbox[t]{.6\linewidth}{the reconstructed original data}}

        \Statex
        \LineCommentContBefore{for each data fragment's storage information $loc\_info$ in $loc\_list$}
        \For{each $loc\_info_f$ in $loc\_list$}
            \LineCommentContAfter{broadcast to all CSPs to return the data fragment; CSPs returns the data fragment if it exists, $None$ if it does not}
            \State $data\_frag_f \gets$ broadcastQuery($loc\_info_f$)
            \LineCommentContBefore{check the response of the primary CSP; reconstruct the data fragment if the check fails}
            \If {not(checkFrag($data\_frag_f[0]$)}
                \LineCommentContAfter{reconstruct the data fragment while referencing secondary CSPs}
                \State  $frag_f \gets$ reconstFrag($data\_frag_f$)
                \LineCommentContBefore{store the reconstructed data fragment at the \emph{new} primary CSP and update the storage information}
                \State  $loc\_info_f[0] \gets$ storeFrag($frag_f, pri\_CSP_{new}$)
            \EndIf
        \EndFor
        \LineCommentContBefore{reconstruct the original data}
        \State  $org\_data \gets$ reconstData($data\_frag$)
        \State return $org\_data$
    \end{algorithmic}
\end{algorithm}
\subsection{Application on existing methods}
A preliminary analysis of the applicability of the proposed framework on various data splitting techniques was conducted. The analysis looked at the adopted security model, measures for privacy, data splitting mechanism, data fragmentation approach, and the use of encryption. 
It was determined that data splitting methods that introduced randomisation into the data fragmentation process will find difficulties due to the resulting uniqueness of data fragments. Also, it may be possible to recreate similar encrypted data fragments using the same encryption parameters but this is not encouraged.
Finally, methods that preserve higher levels of usability in data fragments support key operations of the proposed framework. 
\section{Semantic data splitting leveraging on existing data}\label{sect:sem-data-splitting-leveraging}
The proposed framework was applied to a data splitting mechanism \cite{sanchez2017privacy}. Textual data is split to be outsourced in clear while supporting various operations. This allowed the proposed framework to present itself distinctly when it searches for data fragments in the multi-cloud. 

The data splitting mechanism is aware of contextual information within the data while evaluating semantics against privacy requirements provided by a data owner during fragmentation and achieves \emph{a priori} privacy guarantees by exploiting the $\mathbb{C}$-$sanitisation$ privacy model \cite{sanchez2016c,sanchez2017toward}. 
Data fragments are constructed by dividing and sorting terms that were determined to risk disclosing \emph{identifiable} information before distributing across the multi-cloud. 
It can be applied to individual instances of unstructured data objects (i.e. single documents) unlike privacy models which require a collection of data sets.

\subsection{Semantic data splitting with the proposed framework}
The proposed framework addresses issues identified in \cite{sanchez2017privacy} on two fronts. First, the number of performed 
operations 
is greatly reduced. Terms that were already found stored do not undergo costly operations needed to allocate terms to a suitable data fragment. Their storage information is simply recorded.
The second observes a decrease in the number of data fragments. Thereby reducing the number of storage locations required. The lesser the number of term allocations, the lesser the number of resulting data fragments. Generated data fragments should still comply with privacy requirements.

The proposed framework assumes the same architecture and security model as \cite{sanchez2017privacy}. However, the trusted proxy will broadcast search queries to CSPs and record the storage locations for queries that received positive responses or $None$ otherwise. Terms whose queries returned positive are referred to as \emph{third-party data fragments}. Third-party fragments do not need to be stored nor allocated to a locally created data fragment. Leftover \emph{identifying} terms are stored locally while leftover \emph{quasi-identifying} terms are split into data fragments before being distributed respectively. Storage information will be used when querying or restoring data. 

\subsection{Implementation, evaluation, and analysis}
Experiments were conducted on a virtual Ubuntu 18.04 instance. Articles were sourced from Wikipedia, as in \cite{sanchez2017privacy}, and segmented into paragraphs that are approximately 1KB in size for evaluation. 
Terms were extracted with Rapid Automatic Keyword Extraction (RAKE) and common terms were used to form a database. Web-based information theoretic assessments were performed with the Bing search engine while queries were keyword word searches on the local database. The local database substitutes querying CSPs to contain the experiment such that the relationship between the number of data fragments to the cost of its construction is clearly defined. 

\begin{table}[htbp]
\scriptsize
\caption{Solution comparison fragment generation}
\centering
\begin{center}
\begin{tabular}{|M{0.125\linewidth}|M{0.155\linewidth}|M{0.11\linewidth}|M{0.055\linewidth}|M{0.035\linewidth}|M{0.04\linewidth}|M{0.035\linewidth}|M{0.05\linewidth}|}
\hline
Solution & Privacy model & Strategy & $frag$ & $id$ & $qid$ & $et$ & Time (min) \\
\hline
\multirow{2}{*}{\parbox{\linewidth}{\centering\cite{sanchez2017privacy}}} & \multirow{2}{*}{\parbox{\linewidth}{\centering \emph{(HIV,virus)-sanitisation}}} 
& Unordered & 26 & 4 & 43 & - & 21 \\
\cline{3-8} 
 &  & Ordered & 26 & 5 & 43 & - & 19 \\
\hline
\multirow{2}{*}{\parbox{\linewidth}{\centering Proposed framework}} & \multirow{2}{*}{\parbox{\linewidth}{\centering \emph{(HIV,virus)-sanitisation}}} 
& Unordered & 20 & 2 & 28 & 18 & 11 \\
\cline{3-8} 
 &  & Ordered & 20 & 2 & 27 & 17 & 10 \\
\hline
\end{tabular}
\label{tab:r1}
\end{center}
\end{table}

\subsubsection{Evaluation metrics and observations}
The metrics in \cite{sanchez2017privacy} were adjusted to evaluate and feature quantitative measurements for comparison. They are described as follows:
\begin{itemize}
    \item $frag$: Average number of data fragments produced. Determines the number of cloud storage locations required.
    \item $id$: Average number of \emph{identifiers} discovered.
    \item $qid$: Average number of \emph{quasi-identifiers} discovered to construct the data fragments. A smaller number of \emph{quasi-identifiers} results in lesser data fragments produced.
    \item $et$: Average number of terms found in the local database. 
    \item Time (min): Average time required to split data.
\end{itemize}

Similar splitting strategies were enlisted where 
terms and data fragments were either arranged and evaluated according to the level of information it discloses in increasing or decreasing order (ordered) or not (unordered).

Table~\ref{tab:r1} presents the averaged number of extracted terms and data fragments produced. 
The proposed framework observed lower counts of \emph{identifiers}, \emph{quasi-identifiers}, and data fragments. Directly affecting the number of operations performed during the data splitting process. 
The probability of positive responses to queries increased as the local database already contained terms that were associated with the instantiation of the privacy model. The increased number of terms found in the local database implies a decrease in the number of operations performed to generate data fragments, thus lesser time is required to split data.
Web-based information theoretic assessments may have influenced the data splitting process \cite{Bosch2016EstimatingSE}. 

\subsubsection{Third-party data fragments and queries}
Risks from sourcing data from the multi-cloud are reduced. Information inferred from third-party data fragments cannot be associated with the data owner because they belong to third-party entities. Hence, these data fragments neither need to conform to privacy requirements of the data owner nor are subject to the costly data splitting process.
Additionally, the proxy may split queries in a way that only partially leaks information to prevent CPSs from inferring information. Sub-queries can be sent to CSPs observing their level of sensitivity. Where sub-queries with higher levels of privacy or sensitivity are sent to CSPs with higher levels of privilege or trust as in \cite{dev2012approach}. Data owners may employ techniques such as secret searches to further protect their queries. The authors are aware that data owners also have the ability to infer information from the responses of the CSPs, however, this is outside the scope of this work. This work also discounts the ability to map data fragments to data objects not of their own as it violates privacy. 


\section{Conclusion} \label{sect:con}
The proposed data splitting framework leverages existing data in the multi-cloud to improve data splitting mechanisms. It introduces a simple yet non-trivial query step to the data splitting process. Various factors were considered for the use of third-party data fragments, including concerns on the privacy of queries and management of shared data fragments. 
The application of the proposed framework was also evaluated and was found suitable for most data splitting mechanisms. Furthermore, extensive analysis of one such application shows that the proposed framework provided enhancements. Although, it assumes the availability of similar, if not the same, third-party data fragments while addressing the reliability and availability of data fragments stored in the Cloud.
The authors would like to explore dynamically managing third-party data fragments and protecting split data queries sent to the Cloud in the future.


\printbibliography

\end{document}